\documentclass[3p,times,twocolumn]{elsarticle}

\usepackage{ecrc}

\volume{00}

\firstpage{1}

\journalname{Nuclear Physics B Proceedings Supplement}

\runauth{C. Weiland}

\jid{nuphbp}

\jnltitlelogo{Nuclear Physics B Proceedings Supplement}

\usepackage{amssymb}
\usepackage[UKenglish]{babel}
\usepackage{amsmath}
\usepackage{subfigure}

\biboptions{numbers, compress}

\usepackage[figuresright]{rotating}

\begin{document}

\begin{frontmatter}

%% Title, authors and addresses

%% use the tnoteref command within \title for footnotes;
%% use the tnotetext command for the associated footnote;
%% use the fnref command within \author or \address for footnotes;
%% use the fntext command for the associated footnote;
%% use the corref command within \author for corresponding author footnotes;
%% use the cortext command for the associated footnote;
%% use the ead command for the email address,
%% and the form \ead[url] for the home page:
%%
%% \title{Title\tnoteref{label1}}
%% \tnotetext[label1]{}
%% \author{Name\corref{cor1}\fnref{label2}}
%% \ead{email address}
%% \ead[url]{home page}
%% \fntext[label2]{}
%% \cortext[cor1]{}
%% \address{Address\fnref{label3}}
%% \fntext[label3]{}

\dochead{}
%% Use \dochead if there is an article header, e.g. \dochead{Short communication}

\title{Flavour violating lepton decays in low-scale seesaws}

%% use optional labels to link authors explicitly to addresses:
%% \author[label1,label2]{<author name>}
%% \address[label1]{<address>}
%% \address[label2]{<address>}

\author{C. Weiland}
\ead{cedric.weiland@uam.es}
\address{Departamento de F\'{\i}sica Te\'orica and Instituto de F\'{\i}sica Te\'orica, IFT-UAM/CSIC,\\
Universidad Aut\'onoma de Madrid, Cantoblanco, 28049 Madrid, Spain}

\begin{abstract}

We present the first complete calculation of flavour violating lepton decays taking into account all supersymmetric (SUSY) and non-SUSY contributions in the context of the
supersymmetric inverse seesaw, a specific SUSY low-scale seesaw model. We consider radiative and 3-body lepton decays as well as neutrinoless $\mu-e$ conversion in muonic atoms and perform a full one-loop calculation
in the mass basis. Taking CMSSM-like boundary conditions for the soft SUSY breaking parameters, we find that cancellations between different contributions
are present in several regions, which might reduce the branching ratios by as much as one order of magnitude. This has important consequences when translating current measurements 
into constraints and estimating the reach of future experiments, and justifies the use of a full calculation. We also show that the ratio of different cross-sections
can discriminate between dominant SUSY or non-SUSY contributions.

\end{abstract}

\begin{keyword}
Neutrino Physics \sep Lepton Flavour Violation \sep Inverse Seesaw \sep Supersymmetry

%% keywords here, in the form: keyword \sep keyword

%% MSC codes here, in the form: \MSC code \sep code
%% or \MSC[2008] code \sep code (2000 is the default)

\end{keyword}

\end{frontmatter}

%%
%% Start line numbering here if you want
%%
% \linenumbers

%% main text
\section{Introduction}
\label{Intro}

Since their experimental confirmation two decades ago, neutrino oscillations have been extensively studied and all but two parameters have been precisely measured:
the neutrinos mass ordering and a CP violating phase~\cite{Forero:2014bxa}. This observation constitutes the only signal of new physics observed so far that absolutely calls for
an extension of the Standard Model (SM).
A simple and
attractive possibility to generate neutrino masses and mixing is the Inverse Seesaw (ISS) mechanism~\cite{Mohapatra:1986aw,Mohapatra:1986bd,Bernabeu:1987gr}.
It extends the SM by adding pairs of fermionic singlets with a
seesaw scale close to the electroweak scale and naturally large neutrino Yukawa couplings. However, the SM suffers from other theoretical and observational issues like the absence
of a dark matter candidate or the hierarchy problem. Supersymmetry (SUSY) naturally solves these problems and the supersymmetric inverse seesaw model addresses all these
issues in a common
framework where all the new physics is located around the TeV scale.

While many neutrino mass generating mechanisms lead to the same phenomenology when it comes to neutrino oscillations, they can be distinguished by searching for the effects of the
new particles that they introduce. Those effects can be modified decay chains at high-energy colliders due to the production of the new particles or indirect effects in low-energy
experiments. Charged lepton flavour violating (cLFV) processes are particularly attractive since they are free from Standard Model background and their
cross-sections strongly depend on the model
considered. Besides, there is an intense experimental effort in this field and huge sensitivity improvements are expected in the future, up to five orders of magnitude for neutrinoless 
$\mu - e$ conversion for example~\cite{Abrams:2012er,PRIME}. In this work, we focus on cLFV radiative and 3-body lepton decays as well as 
neutrinoless $\mu-e$ conversion in muonic atoms. Further details, including all formulas, can be found in~\cite{Krauss:2013gya,Abada:2014kba}.

\section{The supersymmetric inverse seesaw model}
\label{SUSYISS}

The supersymmetric inverse seesaw consists of the MSSM extended by three pairs of gauge singlet superfields, $\widehat{\nu}^c_i$ and
$\widehat{X}_i$ ($i=1,2,3$), with opposite lepton number, $-1$ and $+1$, respectively. The superpotential of this model is given by
\begin{equation}
W=  W_\mathrm{MSSM} + \varepsilon_{ab} Y^{ij}_\nu \widehat{\nu}^C_i \widehat{L}^a_j \widehat{H}_u^b+M_{R_{ij}}\widehat{\nu}^C_i\widehat{X}_j+
\frac{1}{2}\mu_{X_{ij}}\widehat{X}_i\widehat{X}_j\,,
\end{equation}
with $W_\mathrm{MSSM}$ the superpotential of the MSSM. The corresponding terms of the soft SUSY
breaking Lagrangian are given by
\begin{align}
-\mathcal{L}^\mathrm{soft}&=-\mathcal{L}_\mathrm{MSSM}^\mathrm{soft} 
         +   \widetilde\nu^{C}_i m^2_{\widetilde \nu^C_{ij}}\widetilde\nu^{C*}_j
         + \widetilde X^{*}_i m^2_{X_{ij}} \widetilde X_j 
         \nonumber\\
      &
     + (T_{\nu}^{ij}  \varepsilon_{ab}
                 \widetilde\nu^C_i \widetilde L^a_j H_u^b +
                B_{M_R}^{ij}  \widetilde\nu^C_i \widetilde X_j \nonumber \\
      &
                +\frac{1}{2}B_{\mu_X}^{ij}  \widetilde X_i \widetilde X_j
      + \widetilde X^{*}_i m^2_{X \nu^C_{ij}} \widetilde \nu^{C}_j
      +\mathrm{h.c.}) \, ,
\end{align}
with $\mathcal{L}_\mathrm{MSSM}^\mathrm{soft} $ the soft SUSY breaking terms of the MSSM.
The only terms that violate lepton number conservation are the Majorana mass term $\mu_{X_{ij}}\widehat{X}_i\widehat{X}_j $ and the last two terms of the soft SUSY breaking
Lagrangian, $B_{\mu_X}^{ij}  \widetilde X_i \widetilde X_j$ and $\widetilde X^{*}_i m^2_{X \nu^C_{ij}} \widetilde \nu^{C}_j$. Thus, taking them to zero increases the symmetry
of the model, making their smallness natural. 

After electroweak symmetry breaking, the  $9\times 9$ neutrino mass matrix is given by
\begin{equation}
 M_{\mathrm{ISS}}=\left(\begin{array}{c c c}
 0 & m_D^T & 0 \\
 m_D & 0 & M_R \\ 
 0 & M_R^T & \mu_X
 \end{array}\right)\,,
\end{equation}
 in the basis $(\nu_L,\, \nu_R^C,\,X)$, where $m_D = \frac{1}{\sqrt{2}} Y_\nu v_u$ and $v_u/\sqrt{2}$ is the
 vacuum expectation value of the up-type Higgs boson. Assuming $\mu_X \ll m_D \ll M_R$, the neutrino mass matrix can be block-diagonalized to give the effective mass
matrix for the light neutrinos~\cite{GonzalezGarcia:1988rw}
\begin{equation}
 M_{\mathrm{light}}\simeq m_D^T {M_R^T}^{-1} \mu_X M_R^{-1} m_D\,,
\label{Mlight}
\end{equation}
while the heavy neutrinos form pseudo-Dirac pairs with masses corresponding
approximately to the eigenvalues of $M_R$. We can see from Eq.~\ref{Mlight} that the lightness of the active neutrinos is directly related to the overall smallness of $\mu$. This decouples
the smallness of the active neutrino masses from the product $M_R^{-1} m_D$ that controls the active-sterile mixing, potentially allowing for large effects in low-energy observables.

\section{cLFV observables and numerical set-up}
\label{ObsSetup}

In this work, we focused on coherent neutrinoless $\mu-e$ conversion in nuclei, the cLFV radiative decays $\ell_\alpha \to \ell_\beta  \gamma$ and the following
3-body decays:
$\ell_\alpha^- \to \ell_\beta^- \ell_\beta^- \ell_\beta^+$, $\ell_\alpha^- \to \ell_\beta^- \ell_\gamma^- \ell_\gamma^+$ and
$\ell_\alpha^- \to \ell_\beta^+ \ell_\gamma^- \ell_\gamma^-$. The analytical formulas for the corresponding decay widths in terms of form factors, as well as the detailed calculation
of these form factors including all possible contributions at the one-loop level in the mass basis can be found in our main article~\cite{Abada:2014kba}.　The current
upper limits on the branching ratio of these processes and the expected sensitivities are given in table~\ref{sensi}.
\begin{table}[tb!]
\centering
\begin{tabular}{|c|c|c|}
\hline
\small cLFV Process &\small Present Bound &\small Future Sensitivity  \\
\hline
    \small $\mu \rightarrow  e \gamma$ & \small $5.7\times 10^{-13}$~\cite{Adam:2013mnn}  & \small $6\times 10^{-14}$~\cite{Baldini:2013ke} \\
    \small$\tau \to e \gamma$ & \small $3.3 \times 10^{-8}$~\cite{Aubert:2009ag}& \small $ \sim3\times10^{-9}$~\cite{Aushev:2010bq}\\
    \small$\tau \to \mu \gamma$ & \small $4.4 \times 10^{-8}$~\cite{Aubert:2009ag}& \small $ \sim3\times10^{-9}$~\cite{Aushev:2010bq} \\
    \small$\mu \rightarrow e e e$ & \small  $1.0 \times 10^{-12}$~\cite{Bellgardt:1987du} & \small  $\sim10^{-16}$~\cite{Blondel:2013ia} \\
    \small$\tau \rightarrow \mu \mu \mu$ & \small $2.1\times10^{-8}$~\cite{Hayasaka:2010np} & \small $\sim 10^{-9}$~\cite{Aushev:2010bq} \\
    \small$\tau^- \rightarrow e^- \mu^+ \mu^-$ & \small  $2.7\times10^{-8}$~\cite{Hayasaka:2010np} & \small $\sim 10^{-9}$~\cite{Aushev:2010bq} \\
    \small$\tau^- \rightarrow \mu^- e^+ e^-$ & \small  $1.8\times10^{-8}$~\cite{Hayasaka:2010np} & \small $\sim 10^{-9}$~\cite{Aushev:2010bq} \\
    \small$\tau \rightarrow e e e$ & \small $2.7\times10^{-8}$~\cite{Hayasaka:2010np} & \small  $\sim 10^{-9}$~\cite{Aushev:2010bq} \\
    \small$\mu^-, \mathrm{Ti} \rightarrow e^-, \mathrm{Ti}$ & \small  $4.3\times 10^{-12}$~\cite{Dohmen:1993mp} & \small $\sim10^{-18}$~\cite{PRIME} \\
    \small$\mu^-, \mathrm{Au} \rightarrow e^-, \mathrm{Au}$ & \small $7\times 10^{-13}$~\cite{Bertl:2006up} & \small \\
    \small$\mu^-, \mathrm{Al} \rightarrow e^-, \mathrm{Al}$ & \small  & \small $10^{-15}-10^{-18}$ \\
    \small$\mu^-, \mathrm{SiC} \rightarrow e^-, \mathrm{SiC}$ & \small  & \small $10^{-14}$~\cite{Aoki:2012zza} \\
\hline
\end{tabular}
\caption{Current experimental bounds and future sensitivities for the cLFV observables considered.}
\label{sensi}
\end{table}

To perform our calculations and the numerical study, we have used the \texttt{FlavorKit} interface~\cite{Porod:2014xia} which is a newly developed tool that allows for the easy
study of flavour observables in models beyond the SM, performing an automated calculation of the form factors and their numerical evaluation. We make contact with
low-energy neutrino data~\cite{Forero:2014bxa} by constructing the neutrino Yukawa coupling with a Casas-Ibarra parametrization~\cite{Casas:2001sr} modified
for the inverse seesaw~\cite{Basso:2012ew,Abada:2012mc}. In our numerical study, we will keep the entries fixed to
\begin{equation}
Y_\nu = 10^{-2} \cdot \begin{pmatrix}
0.0956 & -0.0589 & 0.0348 \\
0.616 & 0.594 & -0.687 \\
0.404 & 1.78 & 1.91
\end{pmatrix}\,,
\end{equation}
which is possible because $\mu_X$ doesn't affect our observables and can always be used to ensure the compatibility with low-energy neutrino data. We use CMSSM-like boundary
conditions at $M_{GUT}$ and 2-loop RGEs that include the entire flavour structure of the model to evolve parameters between the relevant scales. The detailed procedure is described
in our main article~\cite{Abada:2014kba}. This leads to three types of contributions to cLFV at low-energy: neutrino loops, sneutrino loops
and slepton loops with RGE-induced slepton mixing. Since it is possible in the ISS to simultaneously have large $\mathcal{O}(1)$ neutrino Yukawa couplings and a seesaw
scale close to the
electroweak scale, a large enhancement of all three types of contributions with respect the type I seesaw can be expected. In order to simplify the discussion of
our numerical results, the various contributions have been divided in two categories,
SUSY and non-SUSY. It is worth noting that, here, non-SUSY contributions do not simply correspond to the SM contributions but to those of a type~II 2 Higgs doublet model (2HDM).

\section{Numerical results}
\label{Numres}

In the following, we will present our numerical results as functions of the seesaw scale, given by $M_R$, and the soft SUSY parameters taken to be $m_0 = M_{1/2} = -A_0=M_{SUSY}$
in our plots. When they do not vary, these parameters take the standard values given in table~\ref{SUSYinput}.
\begin{table}[t]
\begin{center}
\begin{tabular}{|c|c||c|c|} \hline
$m_0$ & 1 TeV & $M_{1/2}$ & 1 TeV \\
$A_0$ & -1.5 TeV & $M_R$ & 2 TeV \\
$B_{\mu_X}$ & $100\,\mu_X$ & $B_{M_R}$ & $100\,M_R$ \\
$\tan \beta$ & 10 & sign$(\mu)$ & + \\ \hline
\end{tabular}
\end{center}
\caption{Standard values for the input parameters. $M_R$ and $\mu_X$
  are taken diagonal and degenerate.}
\label{SUSYinput} 
\end{table}
While this choice does not always lead to a mass in agreement with CMS and ATLAS measurements for the lightest CP-even Higgs boson, this does not impact the validity of our results.
Indeed, Higgs-mediated contributions are subdominant for $\tan \beta = 10$ and our results exhibit only a mild dependence on $A_0$ which could be adjusted to obtain the proper
Higgs boson mass.

\begin{figure}[t]
\centering
\subfigure{
\includegraphics[width=\linewidth]{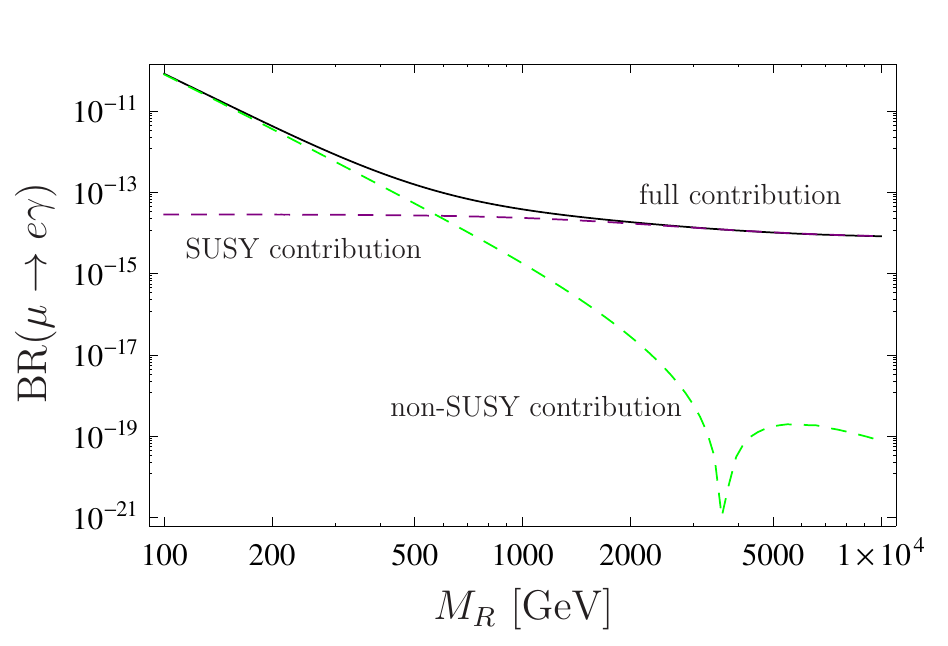}
\label{CRTi-scatter}
}
\subfigure{
\includegraphics[width=\linewidth]{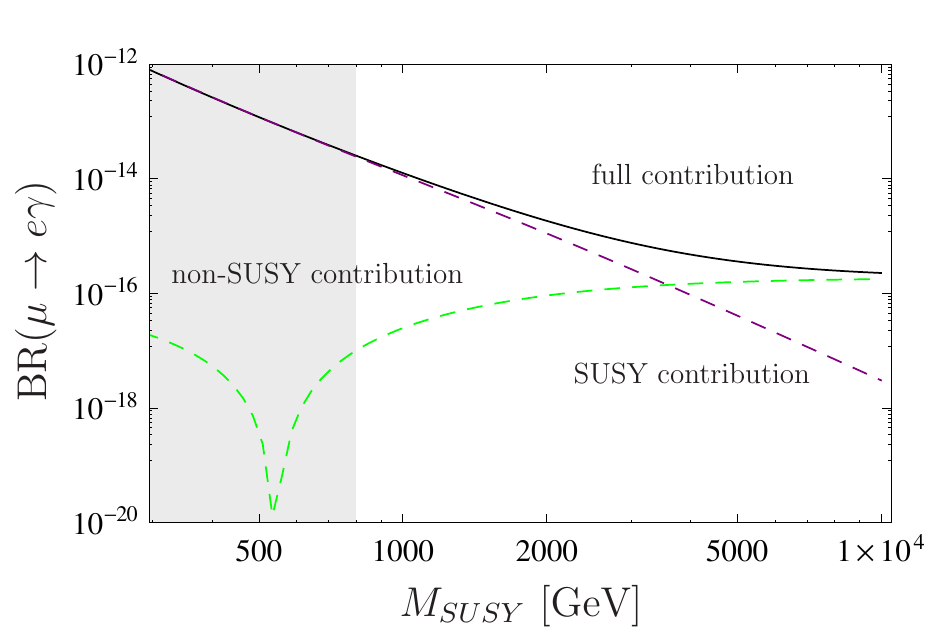}
\label{muegamma-scatter}
}
\caption{$\mathrm{Br}(\mu \rightarrow e \gamma)$ as a function of $M_{SUSY}$ and $M_R$. The grey area is excluded by the ATLAS SUSY search~\cite{Aad:2014wea}.}
\label{CRfig}
\end{figure}

The first observable that we present is the decay $\mu \rightarrow e \gamma$. Its behaviour is representative of other radiative decays and it is one of the most constrained
cLFV processes, with $\mathrm{Br}(\mu \rightarrow e \gamma)< 5.7\times 10^{-13}$ at $90\%$~CL obtained by the MEG experiment~\cite{Adam:2013mnn} and an expected improvement 
in sensitivity to $6\times 10^{-14}$ after upgrade~\cite{Baldini:2013ke}. Our numerical results are shown in Fig.~\ref{CRfig} where the grey area in Fig.~\ref{muegamma-scatter}
corresponds to the region of the CMSSM parameter space excluded by the ATLAS search~\cite{Aad:2014wea}. We can see from both plots that our predictions saturate the current experimental limit 
and the dominant contribution is dictated by the lightest scale, 
$M_R$ or $M_{SUSY}$. If $M_R\ll M_{SUSY}$, the non-SUSY contribution dominates and vice versa. The dip in the non-SUSY contribution comes from a cancellation between
$\nu - H^\pm$ and $\nu - W^\pm$ diagrams, whose matrix elements have opposite signs. The presence of this dip in Fig.~\ref{muegamma-scatter} can be explained by the fact that
scalar masses are functions of $M_{SUSY}$, explaining the dependence of non-SUSY contributions on $M_{SUSY}$. We have explicitly checked that this dependence disappears when taking
the SM limit for the scalar sector.

\begin{figure}[t]
\centering
\subfigure{
\includegraphics[width=\linewidth]{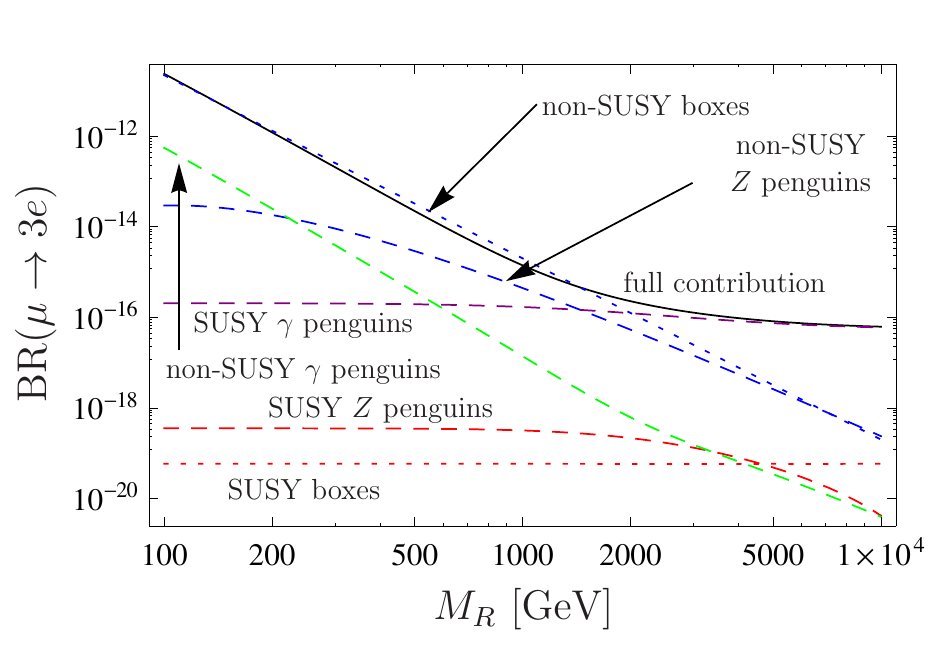}
\label{CRTi-scatter1}
}
\subfigure{
\includegraphics[width=\linewidth]{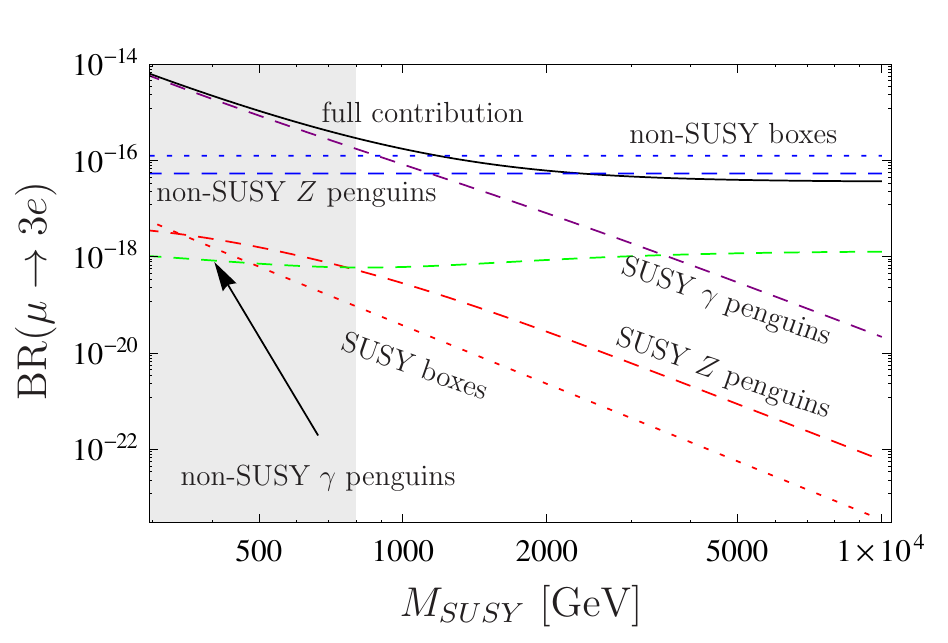}
\label{muegamma-scatter1}
}
\caption{$\mathrm{Br}(\mu \rightarrow 3e)$ as a function of $M_{SUSY}$ and $M_R$. The grey area is excluded by the ATLAS SUSY search~\cite{Aad:2014wea}.}
\label{CRfig1}
\end{figure}

In Fig.~\ref{CRfig1},
we present our results for $\mu \rightarrow 3e$. It is the cLFV 3-body decay with the lowest current experimental limit, the SINDRUM collaboration providing the upper bound
$\mathrm{Br}(\mu \rightarrow 3e)<1.0 \times 10^{-12}$ at $90\%$~CL~\cite{Bellgardt:1987du}. While cLFV radiative decays receive contributions only from dipole operators,
the phenomenology of cLFV 3-body decays is much richer with box diagrams, Z-, $\gamma$- and Higgs-penguins. The latest are not included in Fig.~\ref{CRfig1} since we found their
contribution to be negligible at $\tan \beta = 10$ but it would be sizeable at large $\tan \beta$~\cite{Abada:2011hm}, typically above 50. In the case presented here,
we found that the
dominant SUSY contribution comes from $\gamma$-penguins, while the dominant
non-SUSY contribution
comes from boxes. While the latter can saturate the current experimental bound, it interferes destructively with Z-penguins, which can reduce the total branching ratio
by up to one order of magnitude at large $M_R$. This clearly illustrates the need for a full computation and will have to be taken into account when interpreting future
experimental results.

\begin{figure}[t]
\centering
\subfigure{
\includegraphics[width=\linewidth]{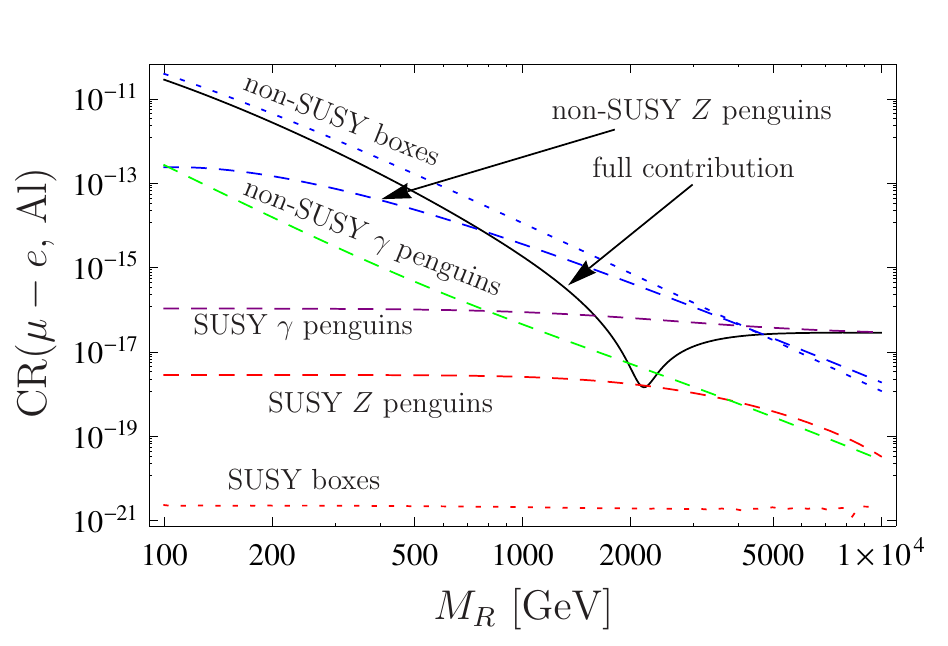}
\label{CRTi-scatter5}
}
\subfigure{
\includegraphics[width=\linewidth]{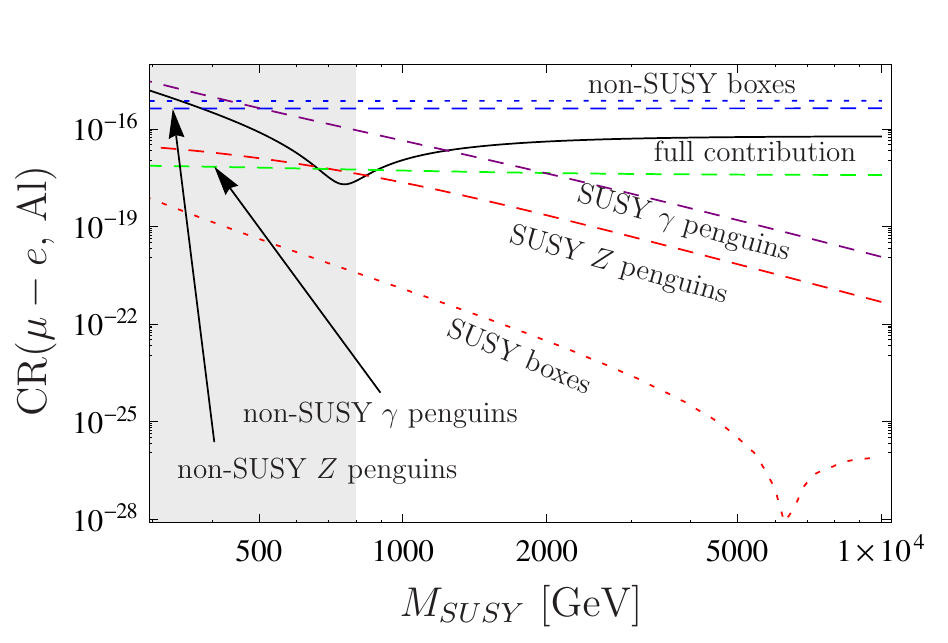}
\label{muegamma-scatter5}
}
\caption{$\mathrm{CR}(\mu - e\,, \mathrm{Al})$ as a function of $M_{SUSY}$ and $M_R$. The grey area is excluded by the ATLAS SUSY search~\cite{Aad:2014wea}.}
\label{CRfig5}
\end{figure}

We can now turn to neutrinoless $\mu -e$ conversion in muonic atoms which similarly exhibits a rich phenomenology as can be seen from Fig.~\ref{CRfig5}.
Here, the cancellation between non-SUSY boxes and Z-penguins is even more important than for $\mu \rightarrow 3e$ and, again, the dominant SUSY contribution comes from $\gamma$-penguin.
Akin to $\mu \rightarrow 3e$, the conversion rate can be large enough to saturate current experimental bounds. The major difference is the appearance of a dip
in the full contribution. This can easily be explained by the fact that the separate (additive)
contributions for the diagrams involving quarks partially have different
relative signs with respect to each other, which was not the case for those involving leptons.

\begin{figure}[t]
\centering
\subfigure{
\includegraphics[width=\linewidth]{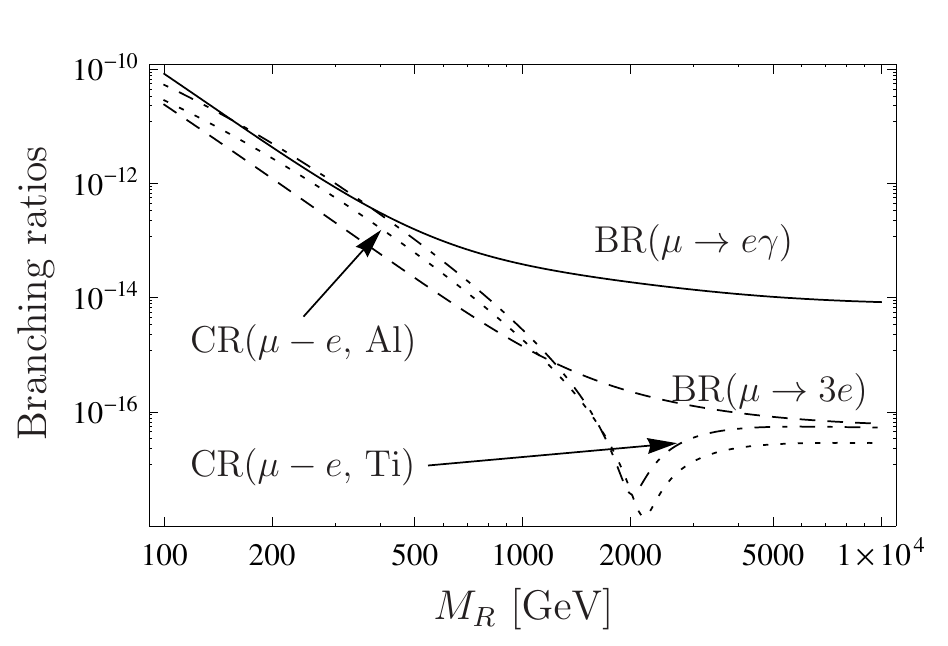}
\label{CRTi-scatter4}
}
\subfigure{
\includegraphics[width=\linewidth]{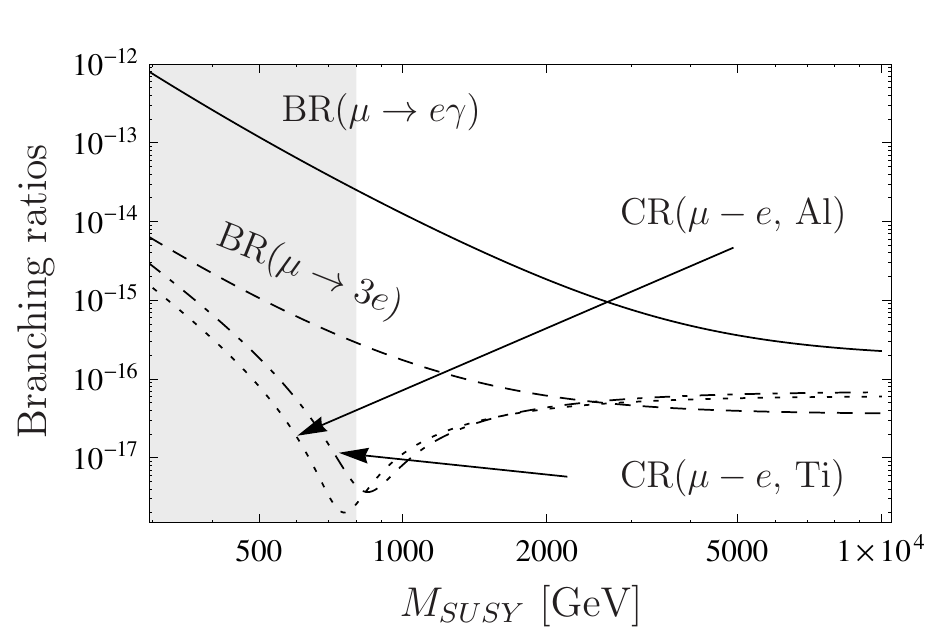}
\label{muegamma-scatter4}
}
\caption{$\mathrm{Br}(\mu\rightarrow e \gamma)$, $\mathrm{Br}(\mu\rightarrow 3 e)$, $\mathrm{CR}(\mu - e\,, \mathrm{Al})$ and $\mathrm{CR}(\mu - e\,, \mathrm{Ti})$
as functions of $M_{SUSY}$ and $M_R$. The grey area is excluded by the ATLAS SUSY search~\cite{Aad:2014wea}.}
\label{CRfig4}
\end{figure}

Having discussed each type of cLFV observables that we considered, let us compare them with the current experimental limits and future sensitivities in mind.
This can be done by plotting
together the three types of observables as in Fig.~\ref{CRfig4}.
First, we can see that the branching ratio of $\mu \rightarrow e \gamma$ is the largest one in most of the parameter space explored. Combined with the very stringent upper limit from
the MEG experiment of $5.7\times 10^{-13}$~\cite{Adam:2013mnn}, this makes $\mu \rightarrow e \gamma$ the most constraining observable in our model nowadays and in the near future.
While the branching ratio of $\mu \rightarrow 3e$ is usually smaller, this observable offers the best mid-term improvement in sensitivity,
the Mu3e proposal~\cite{Blondel:2013ia} mentioning a sensitivity of $\mathrm{Br}(\mu \rightarrow 3e) \sim 10^{-15}$ for the phase I in 2016 and an ultimate sensitivity of $10^{-16}$.
In the long term, experiments searching for neutrinoless $\mu-e$ conversion will provide the largest improvement in sensitivity, up to five orders of magnitude
and reaching sensitivities
down to $10^{-18}$ around 2020. They will then be able to probe a large part of the parameter space, putting strong constraints on this model.

\begin{figure}[t]
\centering
\subfigure{
\includegraphics[width=\linewidth]{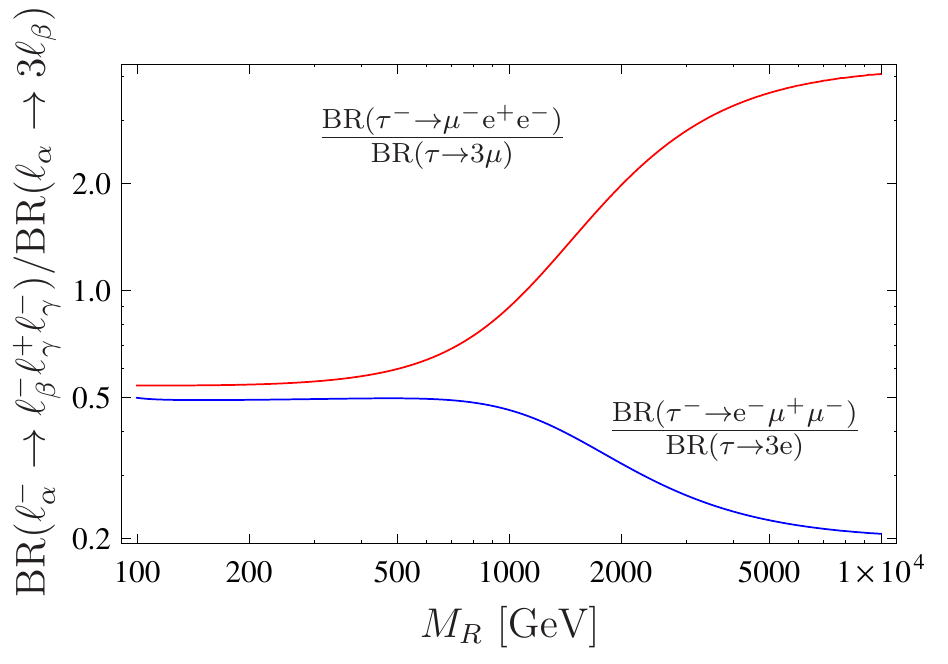}
\label{CRTi-scatter2}
}
\subfigure{
\includegraphics[width=\linewidth]{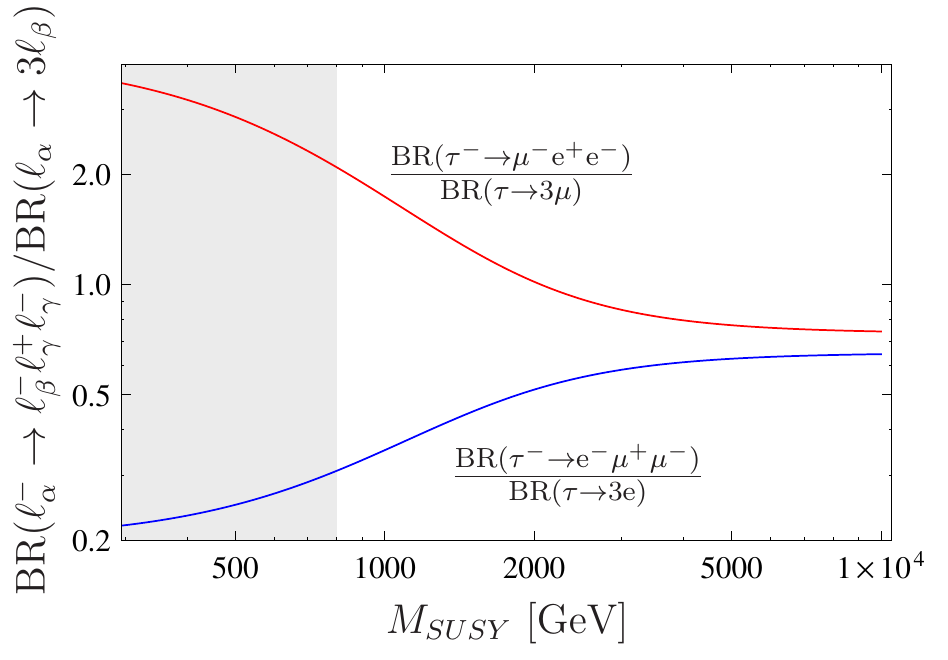}
\label{muegamma-scatter2}
}
\caption{Ratios of 3-body $\tau$ decays as functions of $M_{SUSY}$ and $M_R$. The grey area is excluded by the ATLAS SUSY search~\cite{Aad:2014wea}.}
\label{CRfig2}
\end{figure}

We have focused on $\mu$ decays up to know as they are strongly constrained and they allow to discuss the three types of cLFV observables that we have studied. We refer the interested
reader to our main article~\cite{Abada:2014kba} for results concerning $\tau$ decays. We have also discussed there the impact of the misalignment between light and heavy neutrinos as
well as the effect of a non-degenerate $\mu_X$ matrix. There is however one last observable that would be of great interest at future B factories: the ratio of 3-body
decays $\mathrm{Br}(\tau^- \rightarrow \ell_\beta^- \ell_\gamma^+ \ell_\gamma^-) / \mathrm{Br}(\tau^- \rightarrow 3 \ell_\beta^-)$. A generic prediction of the supersymmetric inverse
seesaw is that $\mathrm{Br}(\tau^- \rightarrow \ell_\beta^- \ell_\gamma^+ \ell_\gamma^-) \simeq \mathrm{Br}(\tau^- \rightarrow 3 \ell_\beta^-)$. However, the exact value of their ratio
is very sensitive to the dominant contribution as can be seen from Fig.~\ref{CRfig2}.
Thus, these ratios can be used to pinpoint the dominant contribution and learn more about the mechanism at the origin of LFV. For completeness,
we note that the branching ratios of $\tau^- \rightarrow \ell_\gamma^- \ell_\beta^+ \ell_\gamma^-$ decays are strongly suppressed, at least by a factor of $10^{-6}$ with
respect to the other 3-body decays, as they require an additional flavour violating vertex.

\section{Conclusion}

In this work, we provide for the first time a complete calculation of cLFV lepton decays and neutrinoless $\mu - e$ conversion in muonic atoms in the supersymmetric
inverse seesaw, taking into
account all SUSY and non-SUSY contributions. We have found that for small $M_R$ or large $M_{SUSY}$, the non-SUSY contributions dominate. In particular, for 3-body decays and $\mu-e$
conversion, the largest contributions come from boxes and Z-penguins which partially cancel each other, reducing the total branching ratio by up to one order of magnitude. In the
large $M_R$ or small $M_{SUSY}$ regime, the SUSY contributions dominate, especially the $\gamma$-penguins. We expect these results to be quite generic and to hold for other
low-scale seesaw models
with right-handed neutrinos and nearly conserved lepton number. In fact, our findings agree with previous results in 
low-scale type I seesaw models~\cite{Ilakovac:2009jf,Alonso:2012ji,Dinh:2012bp,Ilakovac:2012sh}. All types of observables can already be used to constrain the parameter space of the
supersymmetric inverse seesaw. However, due to the huge improvements in experimental sensitivities expected in the upcoming years, the most promising observable depends on the time
scale considered. In the short-term, it will be $\mu \rightarrow e \gamma$, while $\mu \rightarrow 3e$ should be the most constraining around 2016 and $\mu - e$ would give the most
stringent limits around 2020. Finally, we have shown that ratios of 3-body $\tau$ decays will be extremely useful in finding the dominant contribution to cLFV processes.

\section*{Acknowledgements}

C.W.\ receives financial support from the Spanish CICYT through the project FPA2012-31880
and the Spanish MINECO's ``Centro de Excelencia Severo Ochoa'' Programme under grant SEV-2012-0249. C.W.\ also acknowledges partial support
from the European Union FP7 ITN INVISIBLES (Marie Curie Actions, PITN-GA-2011-289442).

\end{document}